\documentclass{ws-ijmpe}

\newcommand{\msun}{{M_{\odot}}}

\newcommand{\kFn}{{k_{F_n}}}
\newcommand{\fmmo}{{\rm fm}^{-1}}
\newcommand{\fmmt}{{\rm fm}^{-3}}
\newcommand{\mev}{{\rm MeV}}
\newcommand{\mevt}{{\rm MeV/fm}^3}
\newcommand{\eos}{EoS~}
\newcommand{\eosp}{EoS}
\newcommand{\eoss}{EoSs~}

\newcommand{\bag}{B^{1/4}}
\newcommand{\gcmt}{{\rm g/cm}^3}

\newcommand{\ergs}{{\rm erg/s}}

\def\rxj1856{\mbox{RX~J1856.5-3754}}
\def\psrotwoofive{\mbox{PSR~0205+6449}}

\begin{document}

\markboth{F.\ Weber et al.}{Ultra-Dense Neutron Star Matter, Strange
Quark Stars, and the Nuclear Equation of State}

\catchline{}{}{}{}{}

\title{Ultra-Dense Neutron Star Matter, Strange Quark Stars, and the
Nuclear Equation of State}

\author{\footnotesize FRIDOLIN
WEBER\footnote{fweber@sciences.sdsu.edu}, ~MATTHEW
MEIXNER,\footnote{meixner@rohan.sdsu.edu}, 
RODRIGO P.\ NEGREIROS\footnote{negreiro@sciences.sdsu.edu}
}

\address{{Department of Physics, San Diego State University\\ 5500
Campanile Drive, San Diego, California 92182-1233, USA}}

\author{MANUEL MALHEIRO\footnote{mane@if.uff.br}}

\address{Instituto de Fisica, Universidade Federal Fluminense\\ 
CEP 24210--340 Niteroi, RJ, Brazil}

\maketitle

%
%
\begin{abstract}
With central densities way above the density of atomic nuclei, neutron
stars contain matter in one of the densest forms found in the
universe. Depending of the density reached in the cores of neutron
stars, they may contain stable phases of exotic matter found nowhere
else in space.  This article gives a brief overview of the phases of
ultra-dense matter predicted to exist deep inside neutron stars and
discusses the equation of state (\eosp) associated with such matter.
\end{abstract}

\section{Introduction}

Neutron stars are dense, neutron-packed remnants of stars that blew
apart in supernova explosions. Many neutron stars form radio pulsars,
emitting radio waves that appear from the Earth to pulse on and off
like a lighthouse beacon as the star rotates at very high
speeds. Neutron stars in x-ray binaries accrete material from a
companion star and flare to life with a burst of x-rays. Measurements
of radio pulsars and neutron stars in x-ray binaries comprise most of
the neutron star observations. Improved data on isolated neutron stars
(e.g. \rxj1856, \psrotwoofive) are now becoming available, and future
investigations at gravitational wave observatories focus on neutron
stars as major potential sources of gravitational waves (see Ref.\
\refcite{villain06:a} for a recent overview). Depending on star mass
and rotational frequency, the matter in the core regions of neutron
stars may be compressed to densities that are up to an order of
magnitude greater than the density of ordinary atomic nuclei. This
extreme compression provides a high-pressure environment in which
numerous subatomic particle processes are believed to compete with
each other\cite{glen97:book,weber99:book}. The most spectacular ones
stretch from the generation of hyperons and baryon resonances
($\Sigma, \Lambda, \Xi, \Delta$) to quark ($u, d, s$) deconfinement to
the formation of boson condensates ($\pi^-$, $K^-$,
H-matter)\cite{glen97:book,weber99:book,heiselberg00:a,lattimer01:a,weber05:a,sedrakian06:a}.
It has also been suggested (strange matter hypothesis) that strange
quark matter may be more stable than ordinary atomic nuclei. In the
latter event, neutron stars could in fact be made of absolutely stable
strange quark matter rather than ordinary hadronic
matter\cite{alcock86:a,alcock88:a,madsen98:b}. Another striking
implication of the strange matter hypothesis is the possible existence
of a new class of white-dwarfs-like strange stars (strange
dwarfs)\cite{glen94:a}. The quark matter in neutron stars, strange
stars, or strange dwarfs ought to be in a color superconducting
state\cite{rajagopal01:a,alford01:a,alford98:a,rapp98+99:a}. This
fascinating possibility has renewed tremendous interest in the physics
of neutron stars and the physics and astrophysics of (strange) quark
matter\cite{weber05:a,rajagopal01:a,alford01:a}.  This paper reviews
the possible phases of ultra-dense nuclear matter expected to exist
deep inside neutron stars and its associated equation of
state\cite{glen97:book,weber99:book,heiselberg00:a,lattimer01:a,weber05:a,sedrakian06:a}.

\section{Neutron Star Masses}\label{sec:masses}

\begin{figure}[tb]
\centerline{\psfig{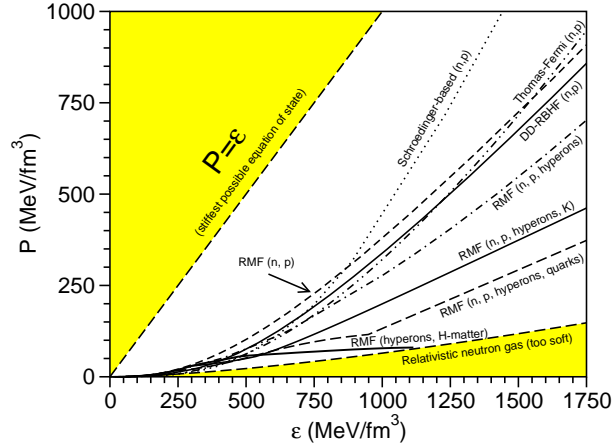}}
\vspace*{8pt}
\caption[]{Models for the \eos (pressure versus energy density) of
neutron star matter\cite{weber05:a}. The notation is as follows:
RMF=relativistic mean-field model; DD-RBHF=density dependent
relativistic Brueckner-Hartree-Fock model; n=neutrons; p=protons;
K=$K^-[u,\bar s]$ meson condensate; quarks=u,d,s; H-matter=H-dibaryon
condensate.}
\label{fig:nucl_eoss} 
\end{figure}

In 1939, Tolman, Oppenheimer and Volkoff performed the first neutron
star calculations, assuming that such objects are entirely made of a
gas of relativistic neutrons\cite{oppenheimer39,tolman39:a}. The \eos
of such a gas is extremely soft (i.e.\ very little additional pressure
is gained with increasing density), as can be seen from Fig.\
\ref{fig:nucl_eoss}, and leads to a
maximum neutron star mass of just $0.7~\msun$. A relativistic neutron
gas thus fails to accommodate neutron stars such as the Hulse-Taylor
pulsar ($M=1.44\, \msun$)\cite{taylor89:a}, and also conflicts with
the average neutron star mass of $1.350 \pm 0.004\, \msun$ derived by
Thorsett and Chakrabarty\cite{thorsett99:a} from observations of radio
pulsar systems.  More than that, recent observations indicate that
neutron star masses may be as high as $\sim 2~ \msun$. Examples of
such very heavy neutron stars are $M_{\rm J0751+1807} = 2.1 \pm 0.2 ~
\msun$\cite{nice05:b}, $M_{\rm 4U\,1636+536} = 2.0 \pm 0.1~
\msun$\cite{barret06:a}, $M_{\rm Vela\, X-1} = 1.86\pm 0.16\,
\msun$\cite{barziv01:a}, $M_{\rm Cyg\, X-2} = 1.78\pm 0.23 \,
\msun$\cite{casares98:a,orosz99:a}.  Large masses have also been
reported for the high-mass x-ray binary 4U\,1700--37 and the compact
object in the low-mass x-ray binary 2S0921--630, $M_{\rm 4U\,1700-37}
= 2.44 \pm 0.27~ \msun$\cite{clark02:a} and $M_{\rm 2S0921-630} = 2.0
- 4.3 \msun$\cite{shahbaz04:a}. respectively. The latter two objects
may be either massive neutron stars or low-mass black holes with
masses slightly higher than the maximum possible neutron star mass of
$~\sim 3 \msun$. This value follows from a general, theoretical
estimate of the maximal possible mass of a stable neutron star as
performed by Rhoades and Ruffini\cite{rhoades74:a} on the basis that
(1) Einstein's theory of general relativity is the correct theory of
gravity, (2) the \eos satisfies both the microscopic stability
condition $\partial P/\partial\epsilon \geq 0$ and the causality
condition $\partial P / \partial\epsilon \leq c^2$, and (3) that the
\eos below some matching density is known. From these assumptions, it
follows that the maximum mass of the equilibrium configuration of a
neutron star cannot be larger than $3.2\,\msun$.  This value increases
to about $5\,\msun$ if one abandons the causality constraint $\partial
P/\partial\epsilon\leq c^2$,\cite{sabbadini77:a,hartle78:a} since it
allows the \eos to behave stiffer at asymptotically high nuclear
densities.  If either one of the two objects 4U\,1700--37 or
2S0921--630 were a black hole, it would confirms the prediction of the
existence of low-mass black holes\cite{brown94:a}. Conversely, if
these objects were massive neutron stars, their high masses would
severely constrain the \eos of nuclear matter.

\section{Composition of Ultra-Dense Neutron Star Matter}

\begin{figure}[tb]
\centerline{\psfig{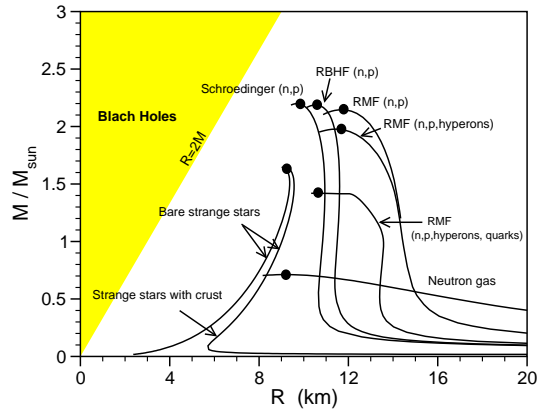}}
\vspace*{8pt}
\caption[]{Mass-radius relationship of neutron stars and strange
stars. The strange stars may be enveloped in a crust of ordinary
nuclear material whose density is below neutron drip
density\cite{alcock86:a,glen92:crust,stejner05:a}. The labels are
explained in Fig.\ \ref{fig:nucl_eoss}. (Figure from Ref.\
\refcite{weber05:a}.)}
\label{fig:MvsR} 
\end{figure}

Models for the \eos of neutron star matter are being computed in
different theoretical frameworks. The most popular ones are the
semi-classical Thomas-Fermi
theory\cite{myers95:a,strobel97:a},
Schroedinger-based treatments (e.g.\ variational approach, Monte Carlo
techniques, hole line expansion (Brueckner theory), coupled cluster
method, Green function
method)\cite{heiselberg00:a,pandharipande79:a,wiringa88:a,akmal98:a},
or relativistic field-theoretical treatments (relativistic mean field
(RMF), Hartree-Fock (RHF), standard Brueckner-Hartree-Fock (RBHF),
density dependent RBHF
(DD-RBHF)\cite{lenske95:a,fuchs95:a,typel99:a,hofmann01:a,niksic02:a,ban04:a},
the Nambu-Jona-Lasinio (NJL) model\cite{lawley06:a,lawley06:b}, and
the chiral SU(3) quark mean field model\cite{wang05:a}). An overview
of \eoss computed for several of these methods is given in Fig.\
\ref{fig:nucl_eoss}. Mass-radius relationships of models of neutron
stars based on these \eoss are shown in Fig.\ \ref{fig:MvsR}.
Any acceptable nuclear many-body calculation must reproduce the bulk
properties of nuclear matter at saturation density and of finite
nuclei. The nuclear matter properties are the binding energy,
$E/A=-16.0$~MeV, effective nucleon mass, $m^*_{\rm N} = 0.79 \,m_{\rm
N}$, incompressibility, $K\simeq 240$~MeV, and the symmetry energy,
$a_{\rm s}= 32.5$~MeV, at a saturation density of $n_0 = 0.16~\fmmt$.

\subsection{Hyperons}

Only in the simplest conception, a neutron star is constituted from
only neutrons. As already discussed in Sect.\ \ref{sec:masses}, this
model fails by far to accommodate observed neutron star masses. At a
more accurate representation, neutron stars will contain neutrons,
$n$, and a small number of protons, $p$, whose charge is balanced by
electrons, $e^-$, according to the reaction $n \rightarrow p +e^- +
\bar\nu_e$ and its inverse.  At the densities that exist in the
interiors of neutron stars, the neutron chemical potential, $\mu^n$,
easily exceeds the mass of the $\Lambda$ so that neutrons would be
replaced with $\Lambda$ hyperons.  From the threshold relation $\mu^n
= \mu^\Lambda$ it follows that this would happen for neutron Fermi
momenta greater than $\kFn \sim 3 \, \fmmo$.\cite{weber05:a}  Such
Fermi momenta correspond to densities of just $\sim 2 n_0$.  Hence, in
addition to nucleons and electrons, neutron stars may be expected to
contain considerable populations of strangeness-carrying $\Lambda$
hyperons, possibly accompanied by smaller populations of the charged
states of the $\Sigma$ and $\Xi$ hyperons\cite{glen85:b}. The total
hyperon population may be as large as 20\% \cite{glen85:b}.

\subsection{Meson condensation}

The condensation of negatively charged mesons in neutron star matter
is favored because such mesons would replace electrons with very high
Fermi momenta. Early estimates predicted the onset of a negatively
charged pion condensate at around $2 n_0$ (see, for instance, Ref.\
\refcite{baym78:a}). However, these estimates are very sensitive to
the strength of the effective nucleon particle-hole repulsion in the
isospin $T=1$, spin $S=1$ channel, described by the Landau
Fermi-liquid parameter $g'$, which tends to suppress the condensation
mechanism. Measurements in nuclei tend to indicate
that the repulsion is too strong to permit condensation in nuclear
matter\cite{barshay73:a,brown88:a}. In the mid 1980s, it was
discovered that the in-medium properties of $K^- [u \bar s]$ mesons
may be such that this meson rather than the $\pi^-$ meson may condense
in neutron star matter\cite{kaplan86:a,brown87:a,lee95:a}.

The condensation is initiated by the schematic reaction $e^-
\rightarrow K^- + \nu_e$.  If this reaction becomes possible in
neutron star matter, it is energetically advantageous to replace the
fermionic electrons with the bosonic $K^-$ mesons. Whether or not this
happens depends on the behavior of the $K^-$ mass, $m^*_{K^-}$, in
neutron star matter.  Experiments which shed light on the properties
of the $K^-$ in nuclear matter have been performed with the Kaon
Spectrometer (KaoS) and the FOPI detector at the heavy-ion synchrotron
SIS at
GSI\cite{barth97:a,senger01:a,sturm01:a,devismes02:a,fuchs06:a}.  An
analysis of the early $K^-$ kinetic energy spectra extracted from
Ni+Ni collisions\cite{barth97:a} showed that the attraction from
nuclear matter would bring the $K^-$ mass down to $m^*_{K^-}\simeq
200~\mev$ at densities $\sim 3\, n_0$. For neutron-rich matter, the
relation $m^*_{K^-} / m_{K^-} \simeq 1 - 0.2 n / n_0$ was
established\cite{li97:a,li97:b,brown97:a}, with $m_K = 495$~MeV the
$K^-$ vacuum mass.  Values of around $m^*_{K^-}\simeq 200~\mev$ may be
reached by the electron chemical potential, $\mu^e$, in neutron star
matter\cite{weber99:book,glen85:b} so that the threshold condition for
the onset of $K^-$ condensation, $\mu^e = m^*_K$ might be fulfilled
for sufficiently dense neutron stars, provided other negatively
charged particles ($\Sigma^-$, $\Delta^-$, $d$ and $s$ quarks) are not
populated first and prevent the electron chemical potential from
increasing monotonically with density.

We also note that $K^-$ condensation allows the conversion reaction $n
\rightarrow p + K^-$. By this conversion the nucleons in the cores of
neutron stars can become half neutrons and half protons, which lowers
the energy per baryon of the matter\cite{brown96:a}. The relative
isospin symmetric composition achieved in this way resembles the one
of atomic nuclei, which are made up of roughly equal numbers of
neutrons and protons.  Neutron stars are therefore referred to, in
this picture, as nucleon stars. The maximum mass of such stars has
been calculated to be around $1.5\,
\msun$\cite{thorsson94:a}. Consequently, the collapsing core of a
supernova, e.g.\ 1987A, if heavier than this value, should go into a
black hole rather than forming a neutron star, as pointed out by Brown
et al.\cite{brown94:a,li97:a,li97:b} This would imply the existence of
a large number of low-mass black holes in our galaxy\cite{brown94:a}.
Thielemann and Hashimoto\cite{thielemann90:a} deduced from the total
amount of ejected $^{56}{\rm Ni}$ in supernova 1987A a neutron star
mass range of $1.43 - 1.52~ \msun$. If the maximum neutron star mass
should indeed be in this range ($\sim 1.5~\msun$), it would pose a
problem to the possible discovery of very heavy neutron stars of masses
around $2~\msun$ (Sect.\ \ref{sec:masses}).
In closing, we mention that meson condensates lead to neutrino
luminosities which are considerably enhanced over those of normal
neutron star matter. This would speed up neutron star cooling
considerably\cite{thorsson94:a,schaab95:a}.

\subsection{H-matter and exotic baryons}

A novel particle that could be of relevance for the composition of
neutron star matter is the H-dibaryon (H=$([ud][ds][su])$), a doubly
strange six-quark composite with spin and isospin zero, and baryon
number two\cite{jaffe77:a}. Since its first prediction in 1977, the
H-dibaryon has been the subject of many theoretical and experimental
studies as a possible candidate for a strongly bound exotic state. In
neutron star matter, which may contain a significant fraction of
$\Lambda$ hyperons, the $\Lambda$'s could combine to form H-dibaryons,
which could give way to the formation of H-dibaryon matter at
densities somewhere above $\sim 4\,
n_0$\cite{tamagaki91:a,sakai97:a,glen98:a}.  If formed in neutron
stars, however, H-matter appears to unstable against compression which
could trigger the conversion of neutron stars into hypothetical
strange stars\cite{sakai97:a,faessler97:a,faessler97:b}.

Another particle, referred to as exotic baryon, of potential relevance
for neutron stars, could be the pentaquark, $\Theta^+ ([ud]^2 \bar
s)$, with a predicted mass of 1540~MeV. The pentaquark, which carries
baryon number one, is a hypothetical subatomic particle consisting of
a group of four quarks and one anti-quark (compared to three quarks in
normal baryons and two in mesons), bound by the strong color-spin
correlation force (attraction between quarks in the color $\bar {\bf
3}_c$ channel) that drives color
superconductivity\cite{jaffe03:a,jaffe05:a}. The pentaquark decays
according to $\Theta^+(1540) \rightarrow K^+ [\bar s u] + n[udd]$ and
thus has the same quantum numbers as the $K^+ n$. The associated
reaction in chemically equilibrated matter would imply $\mu^{\Theta^+}
= \mu^{K^+} + \mu^n$.

\subsection{Quark deconfinement}

It has been suggested already many decades
ago\cite{ivanenko65:a,fritzsch73:a,baym76:a,keister76:a,%
chap77:a,fech78:a,chap77:b} that the nucleons may melt under the
enormous pressure that exists in the cores of neutron stars, creating
a new state of matter know as quark matter. From simple geometrical
considerations it follows that for a characteristic nucleon radius of
$r_N\sim 1$~fm, nucleons may begin to touch each other in nuclear
matter at densities around $(4\pi r^3_N/3)^{-1} \simeq 0.24~\fmmt =
1.5\, n_0$, which is less than twice the density of nuclear
matter. This figure increases to $\sim 11 \, n_0$ for a nucleon radius
of $r_N = 0.5$~fm. One may thus speculate that the hadrons of neutron
star matter begin to dissolve at densities somewhere between around
$2-10\, n_0$, giving way to unconfined quarks.  Depending on
rotational frequency and neutron star mass, densities greater than two
to three times $n_0$ are easily reached in the cores of neutron stars
so that the neutrons and protons in the cores of neutron stars may
indeed be broken up into their quarks
constituents\cite{glen97:book,weber99:book,weber05:a,glen91:pt}. More
than that, since the mass of the strange quark is only $m_s \sim
150$~MeV, high-energetic up and down quarks will readily transform to
strange quarks at about the same density at which up and down quark
deconfinement sets in. Thus, if quark matter exists in the cores of
neutron stars, it should be made of the three lightest quark flavors.
A possible astrophysical signal of quark deconfinement in the cores of
neutron stars was suggested in Ref.\ \refcite{glen97:a}.  The
remaining three quark flavors (charm, top, bottom) are way to massive
to be created in neutron stars. For instance, the creation of charm
quark requires a density greater than $10^{17}\, \gcmt$, which is
$\sim 10^2$ times greater than the density reached in neutron stars.
A stability analysis of stars with a charm quark population reveals
that such objects are unstable against radial oscillations and, thus,
can not exist stably in the universe\cite{weber99:book,weber05:a}. The
same is true for ultra-compact stars with unconfined populations of
top and bottom quarks, since the pulsation eigen-equations are of
Sturm-Liouville type. 

\section{Strange Quark Matter}

It is most intriguing that for strange quark matter made of more than
a few hundred up, down, and strange quarks, the energy of strange
\begin{figure}[tb]
\centerline{\psfig{figure=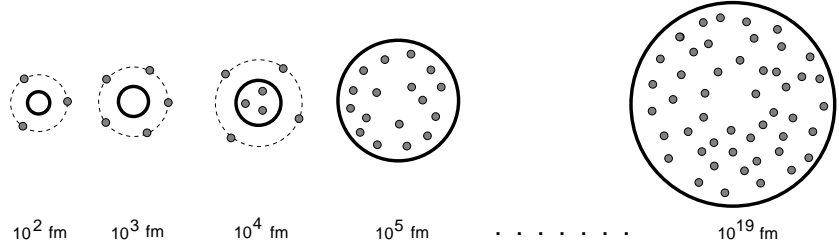,width=11.0cm,angle=0}}
\vspace*{8pt}
\caption[]{Radii of quark bags \cite{giacomelli03:b}.  For masses less
than $10^9$~GeV the electrons (gray dots) are outside the quark bags
(indicated by thick solid circles) and the core+electron system has a
size of $\sim 10^5$~fm. For masses between $10^9$ and $10^{15}$~GeV
the electrons are partially inside the core. For masses greater than
$10^{15}$~GeV all electrons are inside the core, except for a small
electron fraction which forms an electric dipole layer on the surface
of the quark bag (strange star).}
\label{fig:nsizes}
\end{figure} 
quark matter may be well below the energy of nuclear matter, $E/A=
930$~MeV, which gives rise to new and novel classes of strange matter
objects (see Fig.\ \ref{fig:nsizes}), ranging from strangelets at the
low baryon-number end to strange stars at the high baryon number end.
A simple estimate indicates that for strange quark matter $E/A = 4 B
\pi^2/ \mu^3$, so that bag constants of $B=57~\mevt$ (i.e.
$\bag=145$~MeV) and $B=85~\mevt$ ($\bag=160$~MeV) would place the
energy per baryon of such matter at $E/A=829$~MeV and 915~MeV,
respectively, which correspond obviously to strange quark matter which
is absolutely bound with respect to nuclear
matter\cite{madsen88:a,madsen98:b}.

\subsection{Nuclear crust on strange stars}\label{sec:ncrust}

Strange quark matter is expected to be a color superconductor which,
at extremely high densities, should be in the CFL
phase\cite{rajagopal01:a,alford01:a}. This phase is rigorously
electrically neutral with no electrons
required\cite{rajagopal01:b}. For sufficiently large strange quark
masses, however, the low density regime of strange quark matter is
rather expected to form other condensation patterns (e.g.\ 2SC,
CFL-$K^0$, CFL-$K^+$, CFL-$\pi^{0,-}$) in which electrons are
present\cite{rajagopal01:a,alford01:a}. The presence of electrons
causes the formation of an electric dipole layer on the surface of
strange matter, with huge electric fields on the order of
$10^{19}$~V/cm, which enables strange quark matter stars to be
enveloped in nuclear crusts made of ordinary atomic
\begin{figure}[tb]
\centerline{\psfig{figure=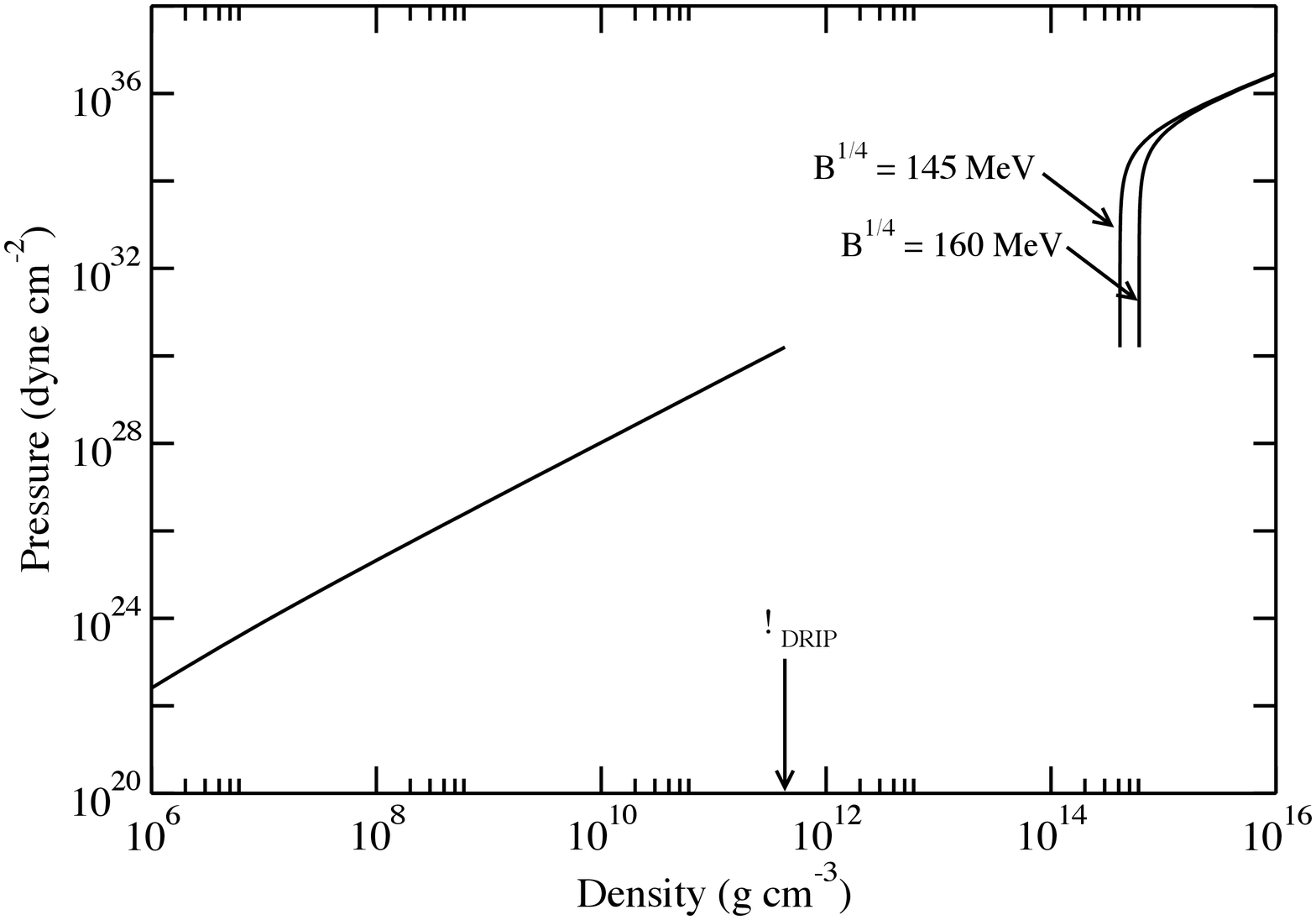,width=7.0cm}}
\vspace*{8pt}
\caption[]{Illustration of the \eos of strange stars with nuclear
crusts (from Ref.\ \refcite{mathews06:a}).}
\label{fig:wdeos} 
\end{figure} 
matter\cite{alcock86:a,alcock88:a,stejner05:a,kettner94:b}.\footnote{Depending
on the surface tension of blobs of strange matter and screening
effects, a heterogeneous crust comprised of blobs of strange quark
matter embedded in an uniform electron background may exist in the
surface region of strange stars\cite{jaikumar05:a}. This heterogeneous
strange star surface would have a negligible electric field which
would make the existence of an ordinary nuclear crust, which requires
a very strong electric field, impossible.}  The maximal possible
density at the base of the crust (inner crust density) is determined
by neutron drip, which occurs at about $4\times 10^{11}~\gcmt$ or
somewhat below\cite{stejner05:a}. The \eos of such a system is shown
in Fig.\ \ref{fig:wdeos}. Sequences of compact strange stars with and
without (bare) nuclear crusts are shown in Fig.\ \ref{fig:MvsR}. Since
the nuclear crust is gravitationally bound to the quark matter core, the
mass-radius relationship of strange stars with crusts resembles the
one of neutron stars and even that of white
dwarfs\cite{glen94:a}. Bare strange stars obey $M \propto R^3$ because
the mass density of quark matter is almost constant inside strange
stars.

\subsection{Strange dwarfs}

For many years only rather vague tests of the theoretical mass-radius
relationship of white dwarfs were possible. Recently the quality and
quantity of observational data on the mass-radius relation of white
dwarfs has been reanalyzed and profoundly improved by the availability
of Hipparcos parallax measurements of several white
dwarfs\cite{provencal98:a}. In that work Hipparcos parallaxes were
used to deduce luminosity radii for 10 white dwarfs in visual binaries
of common proper-motion systems as well as 11 field white
dwarfs. Complementary HST observations have been made to better
determine the spectroscopy for Procyon~B\cite{provencal02:a} and
pulsation of G226-29\cite{kepler00:a}.  Procyon~B at first appeared as
a rather compact star which, however, was later confirmed to lie on
\begin{figure}[tb]
\centerline{\psfig{figure=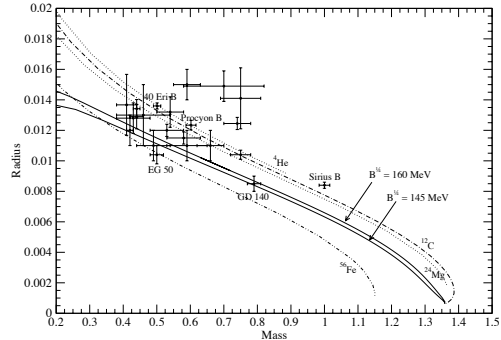,width=7.5cm}}
\caption[]{Comparison of the theoretical mass-radius relationships of
strange dwarfs (solid curves) and normal white
dwarfs\cite{mathews06:a}. Radius and mass are in units of $R_\odot$
and $\msun$, respectively.}
\label{fig:sequences}
\end{figure}
the normal mass-radius relation of white dwarfs.  Stars like Sirius~B
and 40~Erin~B, fall nicely on the expected mass-radius relation too.
Several other stars of this sample (e.g. GD~140, G156--64, EG~21,
EG~50, G181--B5B, GD~279, WD2007--303, G238--44) however appear to be
unusually compact and thus could be strange dwarf candidates
\cite{mathews04:a}.  The situation is graphically summarized in
Fig.\ \ref{fig:sequences}.

\subsection{Surface properties of strange matter}\label{sec:spss}

The electrons surrounding strange quark matter are held to quark
matter electrostatically. Since neither component, electrons and quark
matter, is held in place gravitationally, the Eddington limit to the
luminosity that a static surface may emit does not apply, and thus the
object may have photon luminosities much greater than $10^{38}~\ergs$.
It was shown by Usov\cite{usov98:a} that this value may be exceeded by
many orders of magnitude by the luminosity of $e^+ e^-$ pairs produced
by the Coulomb barrier at the surface of a hot strange star. For a
surface temperature of $\sim 10^{11}$~K, the luminosity in the
outflowing pair plasma was calculated to be as high as $\sim 3 \times
10^{51}~\ergs$.  Such an effect may be a good observational signature
of bare strange stars\cite{usov98:a,usov01:c,usov01:b,cheng03:a}. If
the strange star is enveloped by a nuclear crust however, which is
gravitationally bound to the strange star, the surface made up of
ordinary atomic matter would be subject to the Eddington limit. Hence
the photon emissivity of such a strange star would be the same as for
an ordinary neutron star.  If quark matter at the stellar surface is
in the CFL phase the process of $e^+ e^-$ pair creation at the stellar
quark matter surface may be turned off, since cold CFL quark matter is
electrically neutral so that no electrons are required and none are
admitted inside CFL quark matter\cite{rajagopal01:b}. This may be
different for the early stages of a hot CFL quark star\cite{vogt03:a}.

\section{Proto-Neutron Star Matter}\label{sec:pnsm}

Here we take a brief look at the composition of proto-neutron star
matter. The composition is determined by the requirements of charge
neutrality and equilibrium under the weak processes, $B_1 \rightarrow
B_2 + l + \bar\nu_l$ and $B_2 + l \rightarrow B_1 + \nu_l$, where
$B_1$ and $B_2$ are baryons, and $l$ is a lepton, either an electron
or a muon. For standard neutron star matter, where the neutrinos have
left the system, these two requirements imply that $Q = \sum_{i} q_i
n_{B_i} + \sum_{l=e, \mu} q_l n_l = 0$ (electric charge neutrality)
and $\mu_{B_i} = b_i \mu_n - q_i \mu_l$ (chemical equilibrium), where
$q_{i/l}$ denotes the electric charge density of a given particle, and
$n_{B_i}$ ($n_l$) is the baryon (lepton) number density.  The
subscript $i$ runs over all the baryons considered. The symbol
$\mu_{B_i}$
\begin{figure}[tb]
\begin{center}
\parbox[t]{6.1cm} {\epsfig{file=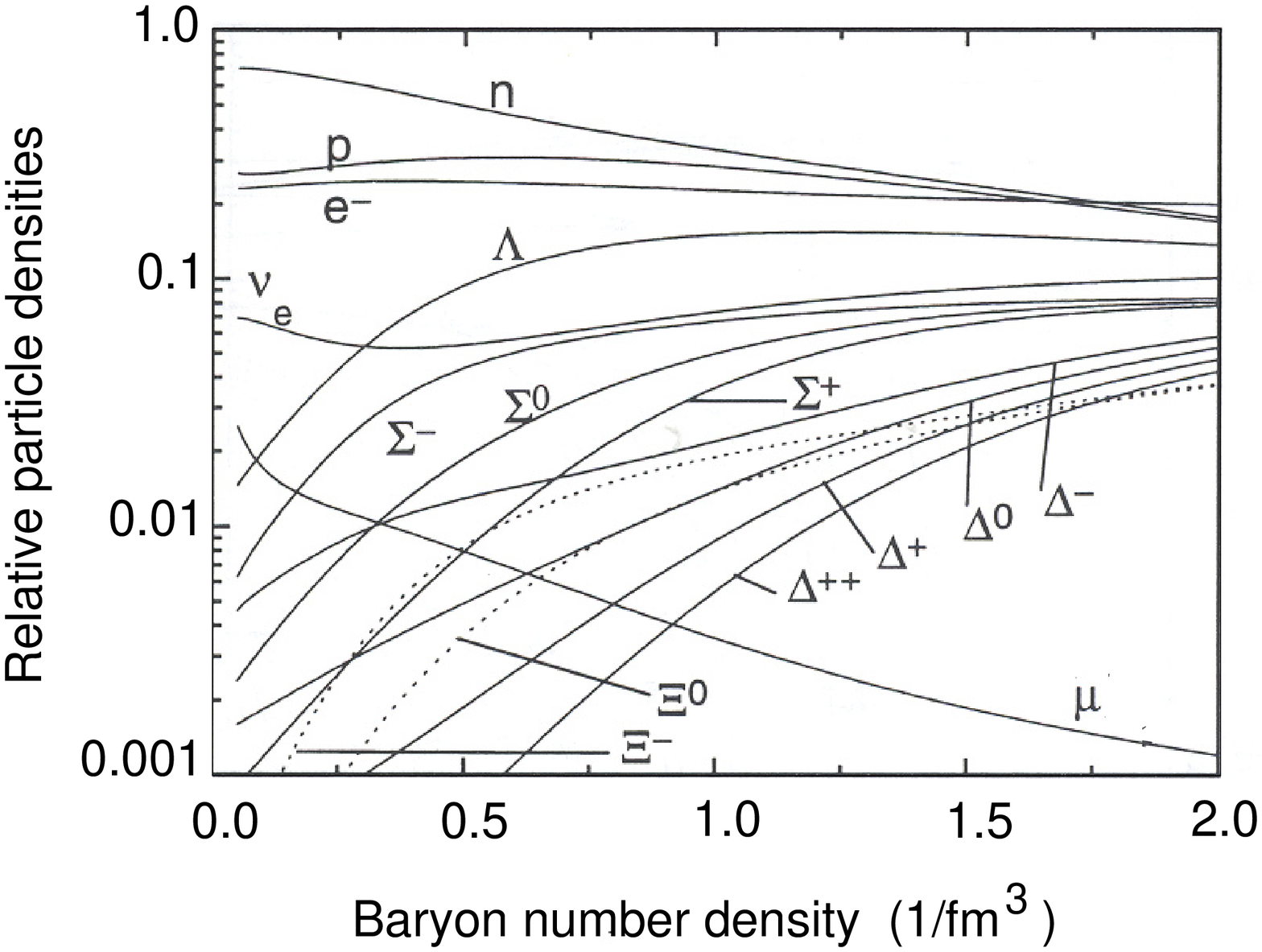,width=6.1cm}
{\caption[]{Composition of hot ($T=40$~MeV) proto-neutron star matter
for $Y_L = 0.3$}
\label{fig:hotpnsm}}}
\ \hskip 0.1cm \
\parbox[t]{6.1cm}
{\epsfig{file=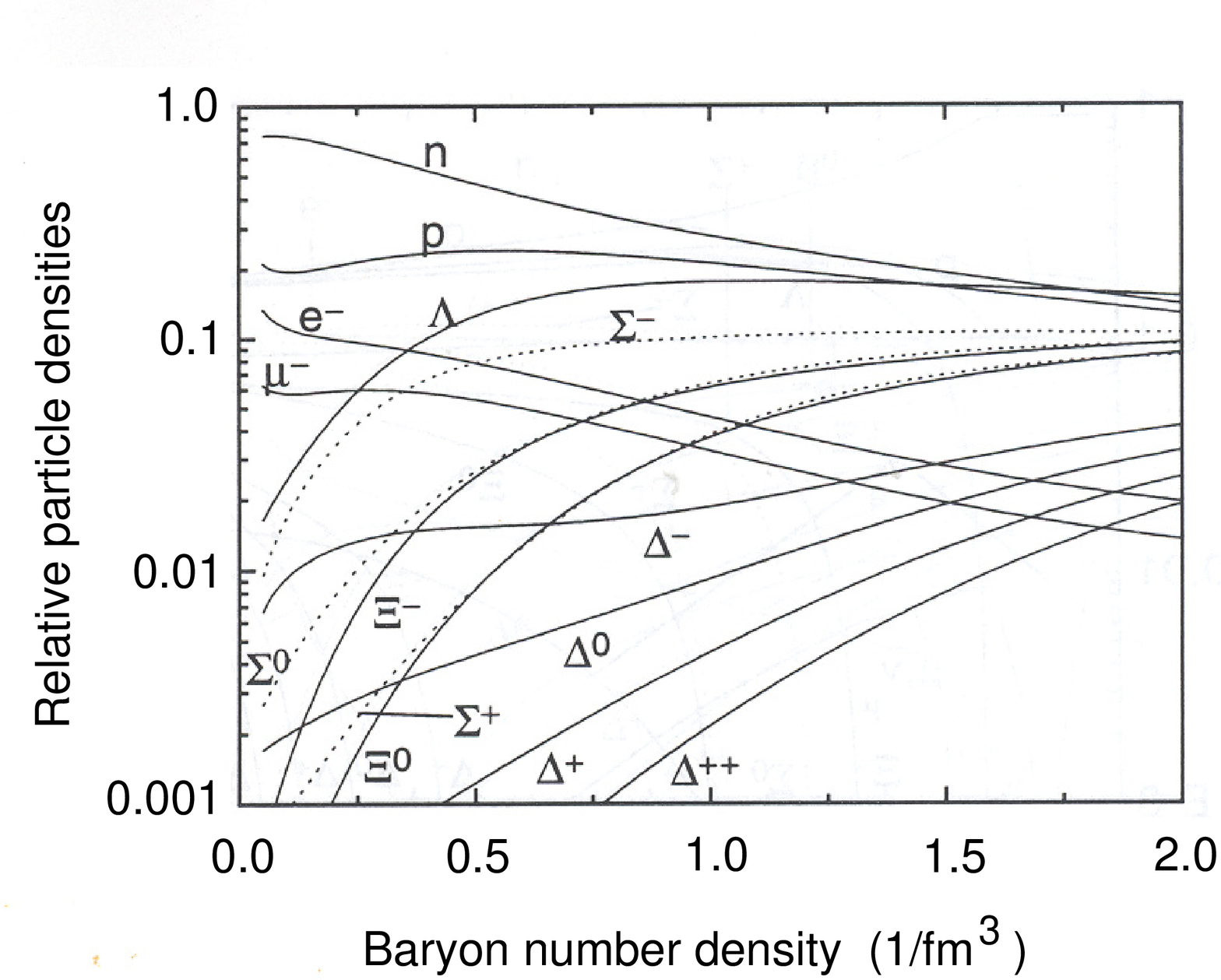,width=6.1cm} {\caption[]{
Same as Fig.\ \ref{fig:hotpnsm}, but for standard neutron star
matter.}
\label{fig:hotnsm}}}
\end{center}
\end{figure}
refers to the chemical potential of baryon $i$, $b_i$ is the
particle's baryon number, and $q_i$ is its charge. The chemical
potential of the neutron is denoted by $\mu_n$.  When the neutrinos
are trapped, as it is the case for proto-neutron star matter, the
chemical equilibrium condition is altered to $\mu_{B_i} = b_i \mu_n -
q_i (\mu_l - \mu_{\nu_l})$ and $\mu_e - \mu_{\nu_e} = \mu_\mu -
\mu_{\nu_\mu}$, where $\mu_{\nu_l}$ is the chemical potential of the
neutrino $\nu_l$. In proto-neutron star matter, the electron lepton
number $Y_L = (n_e+n_{\nu_e})/n_B$ is initially fixed at a value of
around $Y_{L_e} = Y_e + Y_{\nu_e} \simeq 0.3 - 0.4$ as suggested by
gravitational collapse calculations of massive stars. Also, because no
muons are present when neutrinos are trapped, the constraint
$Y_{L_\mu} = Y_\mu + Y_{\nu_mu} =0$ can be imposed. Figures
\ref{fig:hotpnsm} and \ref{fig:hotnsm} show sample compositions of
proto-neutron star matter and standard neutron star matter (no
neutrinos) computed for the RMF approximation. The presence of the
$\Delta$ particle in (proto) neutron star matter at finite temperature
is striking. This particle is generally absent in cold neutron star
matter treated in RMF\cite{glen97:book,weber99:book,glen85:a}.

\section{Electrically Charged Relativistic Star}

As described in Sect.\ \ref{sec:ncrust}, the electric field existing
on the surface of a strange quark star could be as high as
$10^{19}$~V/cm\cite{usov04:a}. The energy density associated with such
a ultra-high field begins to have an impact one the geometry of
space-time, as determined by Einstein's field
\begin{figure}[tb]
\begin{center}
\parbox[t]{6.1cm} 
{\epsfig{file=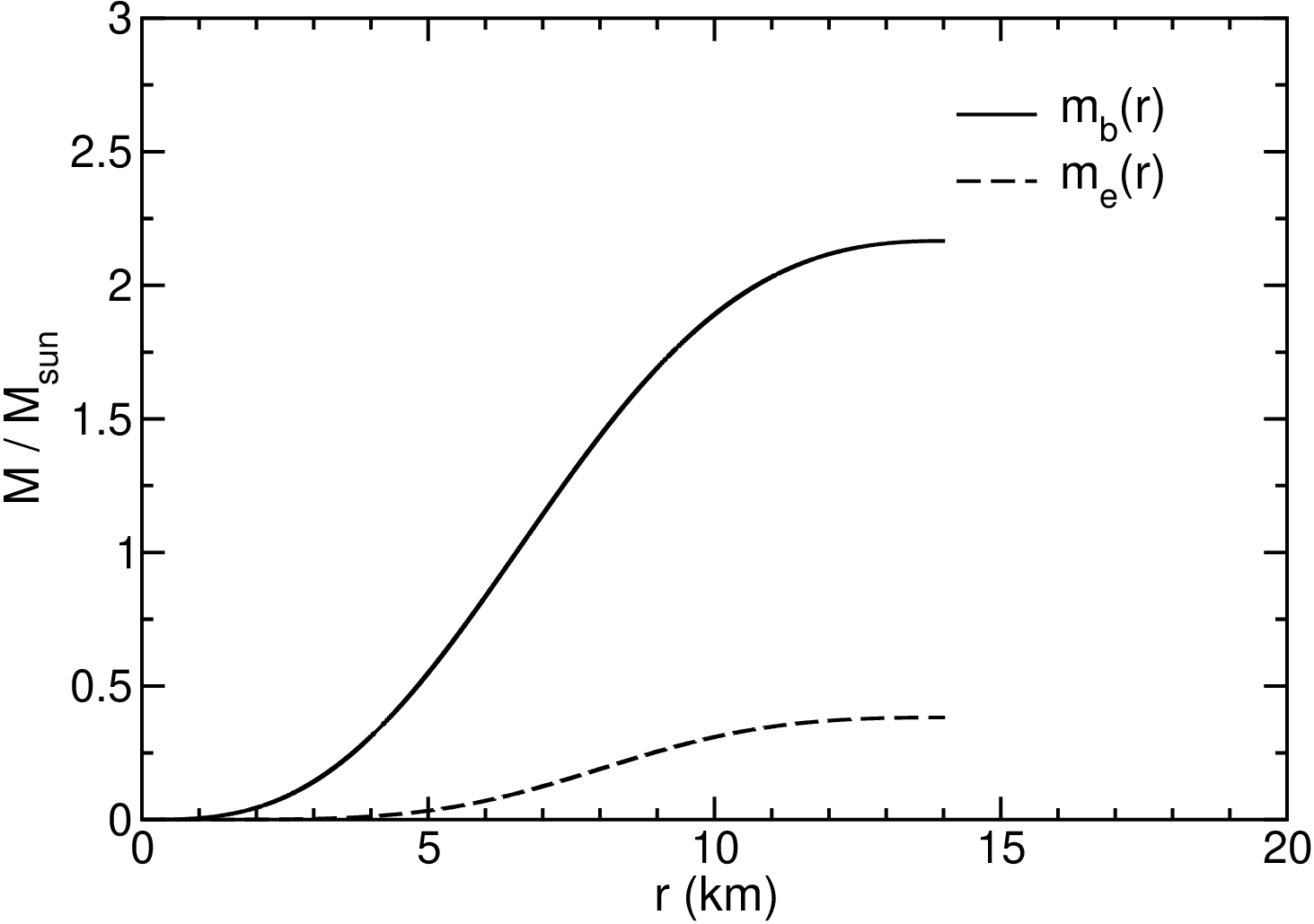,width=6.1cm}
{\caption[]{Star's mass existing as baryonic mass, $m_b$, and electric  
mass-energy, $m_e$, for $f=0.0008$.}
\label{fig:memb_f0008}}}
\ \hskip 0.1cm \
\parbox[t]{6.1cm} 
{\epsfig{file=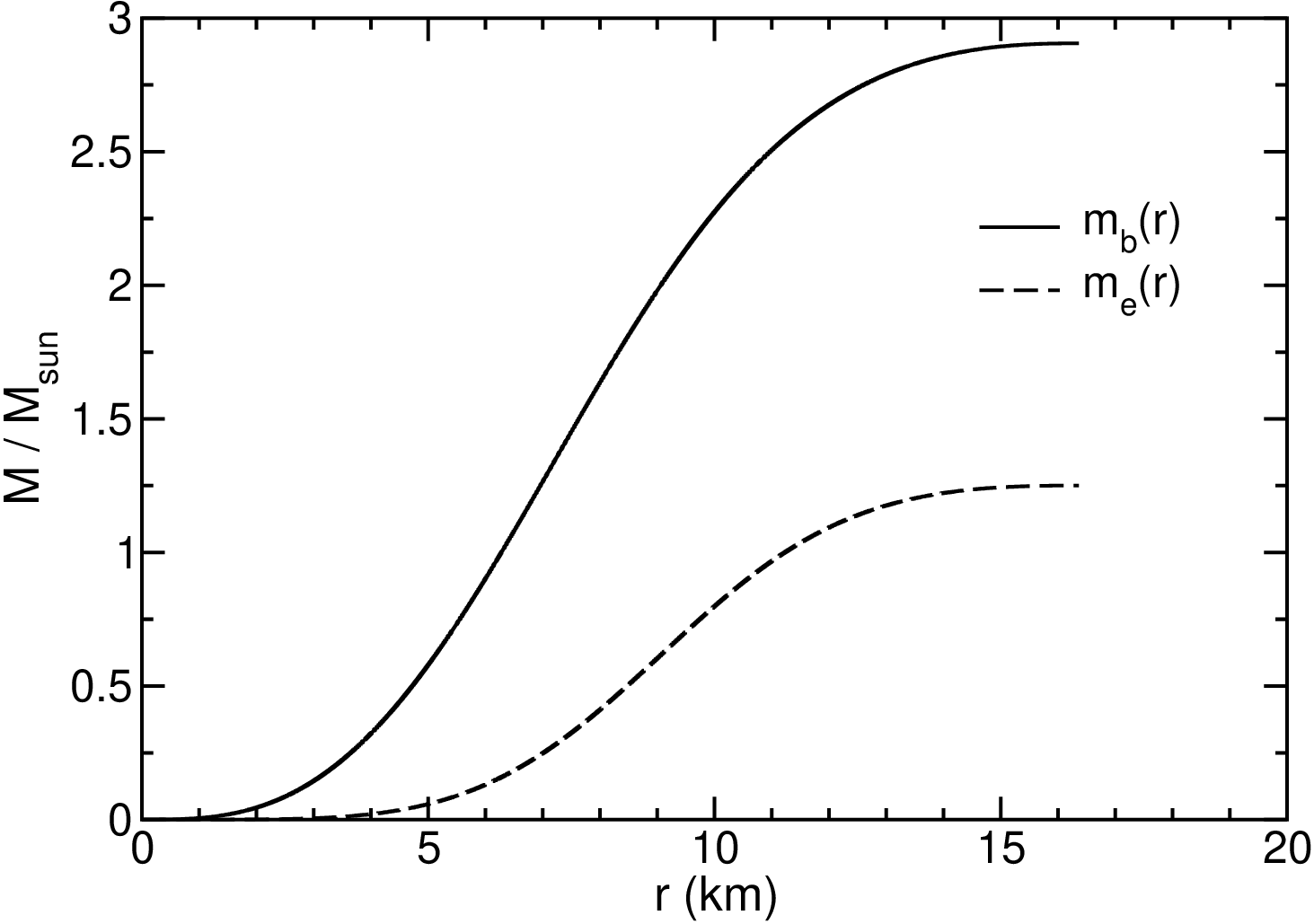,width=6.1cm}
{\caption[]{Same as Fig.\ \ref{fig:memb_f0008}, but for $f=0.001$}
\label{fig:memb_f001}}}
\end{center}
\end{figure}
equation\cite{malheiro04:a,bekenstein71:a,felice95:a}.  In what
follows, we investigate the effects of such ultra-high electric fields
on the properties of relativistic polytropic stars. The consequences
for strange stars are explored in Ref.\ \refcite{negreiros06:a}. 
Since we want to maintain spherical symmetry of the star, the natural
choice for the metric is
\begin{equation}
ds^2=e^{\nu(r)}c^{2}dt^{2}-e^{\lambda(r)}dr^{2}-r^{2}(d\theta^{2}
+\sin^{2}\theta d\phi^{2}) \, . \label{metr}
\end{equation}
The energy-momentum tensor of the star consists of two terms, the
standard term which describes the star's matter as a perfect fluid and
the electromagnetic term,
\begin{equation}
T_{\nu}{}^{\mu} = (p +\rho c^2) u_{\nu} \, u^{\mu} + p \, 
\delta_{\nu}{}^{\mu} + \frac{1}{4\pi} \left( F^{\mu l} F_{\nu l}
+\frac{1}{4 \pi} \, \delta_{\nu}{}^{\mu} \,  F_{kl} \, F^{kl} \right) \, ,
\end{equation}
where the components $F^{\nu \mu}$ satisfy the covariant Maxwell
equations $[(-g)^{1/2} F^{\nu \mu}]_{, \mu}$ = $4\pi J^{\nu}
(-g)^{1/2}$, with $J^\nu$ the four current. For the radial
component one obtains
\begin{equation}
F^{01}(r)= E(r) = e^{-(\nu + \lambda)/2} \, r^{-2} \int_{0}^{r} 4\pi
j^{0} e^{(\nu + \lambda )/2} dr' \, . \label{comp01}
\end{equation}
From last equation we can define the charge of the system as
\begin{equation}
Q(r) = \int_{0}^{r} 4\pi j^{0} r'^{2} e^{(\nu +\lambda)/2} dr'. \label{Q}
\end{equation}
The electric field is then given by
\begin{equation}
E(r) = e^{-(\nu +\mu)/2} \, r^{-2} \, Q(r) \, .
\end{equation}
With the aid of these relations the energy-momentum tensor takes the
form
\begin{equation}
T_{\nu}{}^{\mu} =\left( \begin{array}{cccc}
-\left( \epsilon + \frac{Q^{2} (r)}{8\pi r^4} \right)  & 0 & 0 & 0 \\
0 & p - \frac{Q^{2} (r)}{8\pi r^4 } & 0 & 0 \\
0 & 0 & p + \frac{Q^{2} (r)}{8\pi r^4 }  & 0 \\
0 & 0 & 0 & p  +\frac{Q^{2} (r)}{8\pi r^4 }
\end{array} \right) \, . \label{TEMch}
\end{equation}
Substituting this relation into Einstein's field equation,
$G_{\nu}{}^{\mu} = (8 \pi G/ c^4) T_{\nu}{}^{\mu}$, leads to
\begin{eqnarray}
e^{-\lambda}\left(
-\frac{1}{r^{2}}+\frac{1}{r}\frac{d\lambda}{dr}\right)
+\frac{1}{r^{2}}=\frac{8\pi G}{c^4} \left( p - \frac{Q^{2}(r)}{8\pi
  r^4} \right) \, , \label{fe1q} \\
e^{-\lambda}\left(\frac{1}{r}\frac{d\nu}{dr}+\frac{1}{r^{2}}\right)
-\frac{1}{r^{2}}= - \frac{8 \pi G}{c^4} \left( \epsilon +
\frac{Q^{2}(r)}{8\pi r^4} \right) \, . \label{fe2q}
\end{eqnarray}
The solution for the metric function $\lambda(r)$ is given by
\begin{equation}
e^{-\lambda} = 1 - \frac{Gm(r)}{rc^2} +\frac{GQ^2}{r^2 c^4} \,
. \label{metd}
\end{equation}
The first two terms on the right-hand-side of Eq.\ (\ref{metd})
\begin{table}[tb]
\tbl{Results} {\begin{tabular}{@{}lcclcr@{}} \toprule $f$ & $M~ (M_\odot) 
$ & $M_b~ (M_\odot)$&$M_e~ (M_\odot)$& Ratio & Charge ($ \times 10^{17}$ C) \\
\colrule
0.0    & 1.428 & 1.428  & 0  &   0\%     & 0   \\
0.0001 & 1.439 &  1.436 & 0.00250    &   0.17\% &  260.95  \\
0.0008 & 2.548 &  2.166 & 0.38227    &   15\%   & 3488.01  \\ 
0.001  & 4.156 &  2.905 &   1.250    &   30.07\%& 6727.17  \\ \botrule
\end{tabular} \label{tab:res}}
\end{table}
describe electrically neutral stars, while the third term originates
from the net electric charge distribution inside the star. From Eqs.\
(\ref{fe1q})--(\ref{metd}) one finds that the  mass contained in  a
spherical shell of radius, $m(r)$, is given by
\begin{equation}
\frac{dm(r)}{dr} = \frac{4\pi r^2}{c^{2}} \epsilon +\frac{Q(r)}{c^2
r}\frac{dQ(r)}{dr} \, . \label{dmel}
\end{equation}
The first term on the right-hand-side is the standard result for the
gravitational mass of electrically uncharged stars, while the second
term accounts for the mass change that originates from the electric
field. Finally, the Tolman-Oppenheimer-Vokoff (TOV) equation
generalized to electrically charged stars follows as
\begin{equation}
\frac{dp}{dr} = - \frac{2G\left( m(r) +\frac{4\pi r^3}{c^2} \left( p -
\frac{Q^{2} (r)}{4\pi r^{4} c^{2}} \right) \right)}{c^{2} r^{2} \left(
1 - \frac{2Gm(r)}{c^{2} r} + \frac{G Q^{2}(r)}{r^{2} c^{4}} \right)}
(p +\epsilon) +\frac{Q(r)}{4 \pi r^4}\frac{dQ(r)}{dr} \,
. \label{TOVca}
\end{equation}
To solve the TOV equation, we need to specify the charge distribution
inside the star. Here, we will follow the approach of Ref.\
\refcite{malheiro04:a,ray03:a} and assume that the charge distribution
is proportional to the energy density, $j^0(r) = f \times \epsilon$,
where $f$ is a constant which essentially controls the amount of net
electric charge carried by the star. Moreover, we choose as initial
and boundary conditions of the problem the following values, $\epsilon
(0) = 1550~\mevt$, $Q(0) = m(0) = \lambda(0) = \nu(0) =0$, and $p(R)
=0$.  In this study we focus on the difference between the mass-energy
coming from baryonic matter, here represented by the polytropic \eosp,
and from the electric field. To this aim we rewrite Eq.\ (\ref{dmel})
as ($b$=baryonic matter, $e$=electric field)
\begin{equation}
\frac{dm(r)}{dr} = \frac{dm_b}{dr} + \frac{dm_e}{dr}\, , ~~~ {\rm
where}~~ \frac{dm_b}{dr} = \frac{4\pi r^2}{c^{2}} \epsilon \, , ~~
\frac{dm_e}{dr}= \frac{Q(r)}{c^2r}\frac{dQ(r)}{dr}\, .
\end{equation}
Table~\ref{tab:res} summarizes the properties of several electrically
charged sample stars. 
Figures \ref{fig:memb_f0008} and \ref{fig:memb_f001} show, for two
selected values of $f$, how much of the star's total (gravitational)
mass exists as baryonic mass ($m_b$) and how much mass-energy ($m_e$)
is associated with the electric field energy. For a strongly charged
star, $f=0.001$, we find that 30\% of the star's mass ($1.25~\msun$)
is actually electrostatic energy.

\section*{Acknowledgments}

The research of F.\ Weber is supported by the National Science
Foundation under Grant PHY-0457329, and by the Research Corporation.


\end{document}